\documentclass[12pt, letterpaper]{article}

\usepackage{lineno,hyperref}
\modulolinenumbers[5]

\usepackage{booktabs}
\usepackage{amsmath}
\usepackage{subfig}
\usepackage[inline]{enumitem}
\usepackage{url}

\newcommand{\code}[1]{\texttt{#1}}

\usepackage{authblk}
\usepackage{graphicx}

\title{Automated data validation: an industrial experience report}

\author[1]{Lei Zhang\thanks{Corresponding Author: leizhang@ryerson.ca}}
\author[2]{Sean Howard}
\author[2]{Tom Montpool}
\author[2]{Jessica Moore}
\author[2]{Krittika Mahajan}
\author[1]{Andriy Miranskyy\thanks{Corresponding Author: avm@ryerson.ca}}
\affil[1]{Department of Computer Science, Ryerson University, Toronto, Canada}
\affil[2]{Environics Analytics, Toronto, Canada}

\begin{document}
\maketitle

\begin{abstract}
There has been a massive explosion of data generated by customers and retained by companies in the last decade. However, there is a significant mismatch between the increasing volume of data and the lack of automation methods and tools. The lack of best practices in data science programming may lead to software quality degradation, release schedule slippage, and budget overruns. To mitigate these concerns, we would like to bring software engineering best practices into data science. Specifically, we focus on automated data validation in the data preparation phase of the software development life cycle.

This paper studies a real-world industrial case and applies software engineering best practices to develop an automated test harness called RESTORE. We release RESTORE as an open-source R package. Our experience report, done on the geodemographic data, shows that RESTORE enables efficient and effective detection of errors injected during the data preparation phase. RESTORE also significantly reduced the cost of testing. We hope that the community benefit from the open-source project and the practical advice based on our experience. 
\end{abstract}

\section{Introduction}\label{sec:intro}

Big data and data analytics are revolutionizing modern business processes of sales and marketing. McKinsey Analytics's report shows that big data and data analytics technologies have served as a fundamental technology for artificial intelligence and also enabled a new big data ecosystem~\cite{Analytic52:online}. IDC says that worldwide data volume will grow 61\% to 175~ZB by 2025~\cite{DataAge292:online}. Wikibon predicts that the big data revenues for software and services will reach US \$103~B in 2027 with an annual growth rate of 10.48\%~\cite{10Charts51:online}. 

Besides business opportunities, big data also brings challenges to data scientists. Poor data quality leads to failures of their machine learning models. How to improve the data quality is an active topic in both academia and industry. Data validation is known as an essential process to ensure data quality so that the data are both correct and meaningful for the next step of data analytics. However, due to the large volume, the fast velocity, and the wide variety of generated big data, the data science community is seeking for a more efficient way to ensure data quality. Modern data validation tools have the ability to automate this process to improve the efficiency~\cite{bonter2012data, paygude2013automated, paygude2013automation, rathika2014automated, hynes2017data, polyzotis2017data, polyzotis2019data}. Automation is also one of the most important concepts in continuous integration. However, to the best of our knowledge, automated testing methods and tools are still lacking a mechanism to detect data errors in the datasets, which are updated periodically, by comparing different versions of datasets.

There is another underlying challenge. Data science programmers implementing data analysis algorithms often may not have a computer science or software engineering training as they come from other branches of science, such as mathematics and statistics. Thus, many data science programmers will not have formal training in developing and maintaining complex software solutions. Based on the literature~\cite{Cao17,ZhangZJSXX17} and discussions with practitioners, the programming practices of data science programmers often resemble those of programmers from 1960s that triggered a software crisis leading to the creation of software engineering in 1968~\cite{Wirth08}. Similar issue of lacking best practices also happens in scientific-application software~\cite{Kelly07}. These issues are currently prevalent in the data science community, where existing solutions are of low quality, unmaintainable, and non-evolvable~\cite{Kim0DB16, KimZDB18}. All of this leads to significant economic loss, as the cost of evolving data science products is higher than it should be.

Theoretically, one can try to address these issues by forcing existing rigorous software engineering best practices on data science programmers; realistically, it will (almost surely) work ineffectively~\cite{Kim0DB16, KimZDB18}. This will happen because the programmers lack foundational training and may not have time, interest, or energy to learn laborious practices. Thus, one needs to either tailor existing software engineering techniques or create new lightweight software engineering techniques that would be adopted by data science programmers. Our ultimate \textbf{goal} is to improve the state-of-the-practice and to bring lightweight yet rigorous software engineering into data science.

Automation is crucial for efficient delivery in data-driven software development~\cite{amershi2019software}, especially in two areas: 1)~data preparation and 2)~deployment. In this paper, we focus on the former --- creating automated data validation for verifying the correctness of data preparation and cleansing processes.

In a nutshell, our \textbf{contributions} in this paper are two-fold. First, we report issues that a team of data scientists experience while processing geodemographic\footnote{Geodemography is an area of market research, specializing in profiling economic and demographic characteristics of geographical areas~\cite{goss_1995_geodem}. Geodemographic datasets are typically hierarchical, i.e., tree-like. For example, a group of postal codes belong to a town, the group of towns belong to a county, and a group of counties belong to a province.} datasets. The geodemographic data have to be updated periodically (e.g., due to arrival of new census data). Such an update entails data preparation and cleansing, which is often error-prone. 

Second, to combat the issues experienced during the data update, we present an automated data validation framework called RESTORE. This framework automates the statistical tests introduced by the team as part of the generation of best practices. RESTORE compares the previous version/vintage\footnote{From hereon, we will use the term \textit{version} and \textit{vintage} interchangeably.} of a dataset with a new one and reports potential issues. The framework is written in R language~\cite{r_core} and is released as an open-source R package on GitHub~\cite{restore_repo}.\footnote{\url{https://github.com/miranska/restore}}

By adopting the RESTORE R package in an industrial setting, we seek answer to the question~--- \textbf{does the RESTORE package improve the efficiency of the data validation procedure, i.e., identifying data errors with less time and human resources?}

RESTORE introduces a set of test cases that our team found useful empirically. However, this set is not exhaustive. Thus, we designed RESTORE so that it can be easily extended with additional test cases, if other data science teams decide to introduce new test cases that would help them in validating their data. We welcome contributions to the code via GitHub. Moreover, the team come up with a set of best practices based on our experiences, and we hope data science practitioners who are facing data validation challenges benefit from our take-away messages.

The rest of the paper is structured as follows. In Section~\ref{sec:practice}, we introduce the background of data validation, geodemographic data, and our industrial settings. Section~\ref{sec:solution} discusses our proposed method for automated data validation. Section~\ref{sec:tests} presents the details of data tests. In Section~\ref{sec:discussion}, we analyze the pros, cons, and assumptions of our method. Section~\ref{sec:restore} introduces the interface of RESTORE, our validation, threats to validity, potential extensions of RESTORE, and take-away messages. Section~\ref{sec:related} depicts related work. Finally, Section~\ref{sec:conclusion} concludes the paper.

\section{Background}\label{sec:practice}

In this section, we first discuss the importance of data validation (Section~\ref{sec:importance_validation}). Then, we review the existing solutions (Section~\ref{sec:existing_solutions}). We also give an introduction of the characteristics of geodemographic data (Section~\ref{sec:background}). Finally, we illustrate the data development process in our industrial settings (Section~\ref{sec:ea}). 

\subsection{Importance of data validation}\label{sec:importance_validation}

Data-driven software development consists of both data-oriented stages (e.g., data collection and data cleaning) and model-oriented stages (e.g., model training and model evaluation)~\cite{amershi2019software}. The first stage is often data collection where data scientists look for and integrate available datasets or collect their own. 

A typical data analytics company\footnote{Data analytics companies provide data analytics services by analyzing the acquired data to assist businesses such as software development, market analysis, improving operational efficiency, etc.} will have to collect multiple datasets at this stage, many of which will be updated regularly. For example, a ledger of trades from a foreign office of a bank may be updated daily, macroeconomic metrics data --- quarterly, and census data --- yearly. 
During an update, values of variables are typically refreshed. In addition, observations and variables can be added or removed in the new version of the dataset. 
Once updated data are received, the data development team will clean and transform the data. Then the data will be passed to software development and testing teams that will ingest the data into data science software products that customers will use to access and analyze the data.

\begin{figure}
    \centering
    \includegraphics[width=\textwidth]{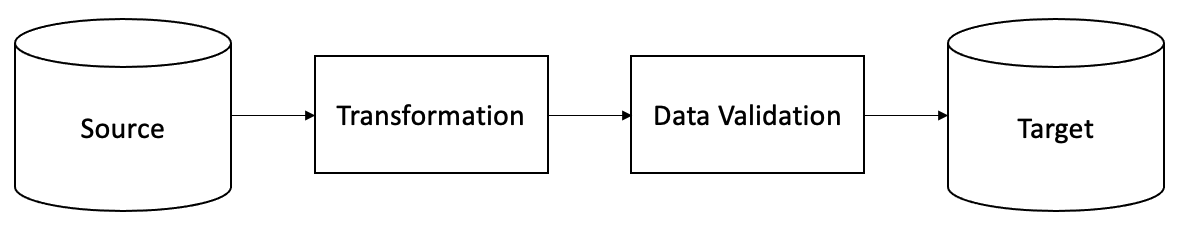}
    \caption{Data validation in ETL pipeline}
    \label{fig:etl}
\end{figure}

The scenario above is a typical Extract, Transform, and Load (ETL) process. Our proposed data validation technique is feasible in any cases where ETL is required, such as machine learning or business analysis. The sequence may vary on a case-by-case basis, but the abstract principle remains the same. Figure~\ref{fig:etl} shows how data validation techniques can be incorporated into the ETL pipeline.

In ETL processes, data transformation routines (manual and automatic) may be defective, leading to errors in the resulting output. Then, the defective output will be used as input data for analytical models, leading to errors in data analytics, especially in machine learning models where input data errors can be amplified over a feedback loop. Such triggers may lead to various failures on the software side, ranging from a failed query in the analytics system (which is easy to detect) to incorrect results (which may be harder to expose). For example, suppose a report suggests that the overall population of Canada is 100 million. In that case, it is an apparent data defect that is easy to spot because the correct answer (as per the 2016 Census) is 36.29 million. However, suppose the report tells us that the population of Canada is 37 million. In that case, a tester may need to verify the numbers manually and must have some context to assess the data appropriately. 

As discussed, data validation is the last step of data development before loading the data to the target platform, which is a counterpart of validation testing in software testing. For example, bugs of internal scripts used for ETL can generate data defects. However, such data defects cannot be revealed by testing the software code but can be reported by validating the data. 
Below, we list some of the root causes that can lead the input data to incorrectness.
\begin{enumerate}
    \item The raw data may be corrupted, e.g., due to incorrect extraction from an external source. 
    \item Data scientists may perform some data transformations manually, which injects some errors. 
    \item The ETL scripts may contain hard-coded values, which worked for the previous version of the dataset but not for the new version. To better illustrate this case, suppose that we have an ``if'' block, i.e., ``if $x \neq 10$'', which works for values $< 10$, but the same ``if'' block can trigger an error in the new version where $x = 11$.
\end{enumerate}
    
Here, we pick two concrete examples from our demos in RESTORE, which can be found in the file of \code{analysis\_results\_hierarchy.xlsx}~\cite{restore_repo}, to illustrate what data defects are and how we can detect them. In the first example, we show a synthetic case where the number of observations changes significantly from the old vintage of the dataset to the new vintage of the dataset (i.e., 16,085 versus 16,018 is reasonable rather than 16,085 versus 160, which is differed by two orders of magnitude). Such a data defect can be detected by magnitude test, i.e., any differences greater than one order of magnitude should be automatically detected and reported to data scientists for further investigation. In the second example, to assess monotonic relationships, we perform the Spearman rank-order correlation test on two variables, and we set a threshold, which is 0.8. As a result, any correlation coefficient with a value of less than 0.8 will be added to the ``alert'' list and manually evaluated by data scientists. 

If no anomalies are revealed during the data preparation stage, these anomalies may pass on to the software developers or testers, leading to failures in system tests. The defective data will be returned to the data development team for a fix. The data development team will now have to fix a defect, then test and reload the data. These delays lead to schedule slippage and budget overruns. Therefore, we need to find a solution to identify such defects as early as possible, saving time and money. In this paper, our goal is to detect data-related defects automatically, streamline data update schedules, and reduce the cost of detecting and fixing the defects. 

\subsection{Existing solutions}\label{sec:existing_solutions}
There are many test frameworks for testing database engines and business logic that alters the data in the databases~\cite{kapfhammer2003family,maule2008impact,haraty2002regression,nanda2011regression,haftmann2005efficient}. Some automated data validation methods for data migration have been proposed~\cite{paygude2013automated, paygude2013automation, rathika2014automated}. In addition, some automated database testing frameworks~\cite{DbFit,DbUnit,NDbUnit,DBTest,SQL} are developed to make sure that the previously captured analytics SQL queries execute successfully in the current version. 
It is important to monitor the quality of data fed to analytics or machine learning models because errors in the input data can nullify the accuracy. Thus, data validation frameworks for data analytics, especially for machine learning models recently, are proposed~\cite{gao2016big, bonter2012data, sadiq2004data, polyzotis2017data, polyzotis2019data, hynes2017data}.

We are not interested in database testing systems. The problem we aim to solve is complementary. We focus on data validation at an early stage --- before a dataset is loaded into the database. More specifically, we are interested in validating a new dataset by comparing it against the predecessor (like regression testing in software engineering). To reach our goal, we need to examine the changes in the data preparation step and flag the erroneous records and variables. 

\subsection{Geodemographic data}\label{sec:background}

As mentioned in Section~\ref{sec:intro}, geodemographic datasets have two characteristics. First, geodemographic datasets are typically hierarchical, e.g., a group of postal codes belong to a town, a group of towns belong to a county, and so on. Second, geodemographic data are usually updated periodically, e.g., due to arrival of new census data.

As an example of periodically updated datasets, many demographic data and surveys (leveraged in geodemography) are updated annually by national census organizations or primary research companies. These data, in turn, get ingested by companies around the globe to improve business decisions. The data ingestion process leads to dataset refreshment and/or transformation, which may introduce new data flaws (e.g., incorrect values) or defects (e.g., changed/missing attributes) in the software. 

The consequences of such errors vary. For example, if the transformed dataset does not have a required variable, the software doing data analysis on this transformed data may fail as it would be unable to find the variable. Such an error would be detected fairly early in the testing of software systems. However, an error may be more subtle: all the variables would be present, but the values of these variables are incorrect, leading to incorrect results generated by the business intelligence software. In one example, a sample report may suggest that the average price of a house for area B is $\$50$K while for area C it is $\$10$M. Both numbers are extreme, but not outside of the realm of possibility. Thus, an analyst may need to manually verify both numbers and must have some context to appropriately assess the resulting numbers. This manual verification is arduous.

Note that our proposed solution --- RESTORE --- is designed to deal with hierarchical data, but it can be applied to ``flat'' datasets too by setting hierarchical depth to zero. Therefore, we hope that that this framework will be helpful to any data scientist dealing with flat or hierarchical data that requires periodic update. 

\subsection{Environics Analytics}\label{sec:ea}

It is important to understand how the data are developed and tested in a typical data analytics/data science project. In this research project, we provide a concrete example to illustrate a typical procedure of data development and testing, based on the process adopted by Environics Analytics, Toronto, Canada (abbreviated to EA in this paper). EA specializes in geodemography and marketing analytics and builds standard and custom data-driven solutions for their clients. Many of their data and services are provided using a Software-as-a-Service (SaaS) approach~\cite{mell_2011_nist}. These services require datasets to be updated multiple times per year. During the update, anomalies in new vintages of datasets may introduce defects in services. The quality assurance team used to manually test the datasets to detect and eliminate the defects, but this process was time- and human-resource-consuming.

EA's SaaS platform hosts data built and maintained by EA, as well as data supplied by EA's partners or clients. Data supplied by partners or clients can be in a variety of formats. Their most common data structure is a tabular dataset that consists of various levels of geographic information. This data architecture has its advantages because it allows for the necessary flexibility to work with different types and sizes of data. In this paper, we assume that all datasets are in tabular format (or can be converted to this format).

Before an automated data testing tool is adopted, the dataset development process in EA is as follows.

\begin{enumerate}
    \item The data team creates a dataset (either a new one or an update of an existing one).
    \item This dataset is loaded into a staging database by the data team.
    \item The software development and quality assurance/testing teams execute a mixture of automated and manual test workloads (against the application) mimicking customers' behaviour (e.g., select a particular geographic area and then run a house prices report). Under the hood, the software layer issues analytic (read-only) queries to the staging database. As part of the software testing, data in analytics reports are assessed, resulting in possible data errors to be uncovered.
    \item If failures (such as those discussed in Section~\ref{sec:intro}) in the analytics reports are observed during the execution of the workloads, a bug report is issued for further investigation. Data bugs may exist in various forms: e.g., errors in raw data, errors in the calculation of ``constructed'' variables as part of the load into the staging database, and errors generated by how the application handles the data.
\end{enumerate}

Once the data team fixes the defects, this team recreates the dataset as necessary, reloads it into the staging database, and hands it over to the software development team for testing (basically, rerunning the above process). This process repeats itself until all the data-related defects are eliminated. Then the dataset is loaded into a production database, and the product is made available to a customer. As mentioned in Section~\ref{sec:intro}, the process is time-consuming and may significantly delay the release (from days to weeks) of the product to customers.

\section{Our solution}\label{sec:solution}

We develop a theoretical foundation for comparison of various versions of the datasets. Rather than performing the exact comparison, we explore methods needed to compare distributions of the data. For example, we compare distributions of the data in a pair of releases (e.g., using Kolmogorov-Smirnov test~\cite{smirnov1939estimation} or Mann-Whitney U test~\cite{mann1947test}). We implement these theoretical findings in an automated regression testing framework. Regression testing is a type of testing, which ensures that the existing functionality of a software product is not broken with new changes, i.e., the functionality does not regress~\cite{huizinga2007automated,lewis2000software}. In our case, the functionality is data-centric. 

The framework enables automatic testing of datasets immediately after generation of a new version of the data, thus, ensuring that defects in the data are captured early in the process, before the dataset is shipped off by the data development team. We do not alter the original inputs (while following best practices of functional programming). Instead, we take a sequence of transformations of datasets and analyze the results of the transformation. The optimal goal of this study is to shorten the development and testing cycle, reducing the probability of schedule slippage and freeing resources to focus on more complex workloads, thus, 

\begin{enumerate}
    \item Improving overall product quality (as teams will have more time and resources to identify complex defects that otherwise would be ``masked'' by simpler defects, which can be caught by the automated regression testing~\cite{vanMegen_1995,rafi2012}), and
    \item Reducing development costs (the savings will manifest themselves because the cost of creating, maintaining, and executing test cases will be lower than the cost of manual testing)~\cite{dustin1999automated}.
\end{enumerate}

\begin{figure}[t]
    \centering
    \includegraphics[width=\columnwidth]{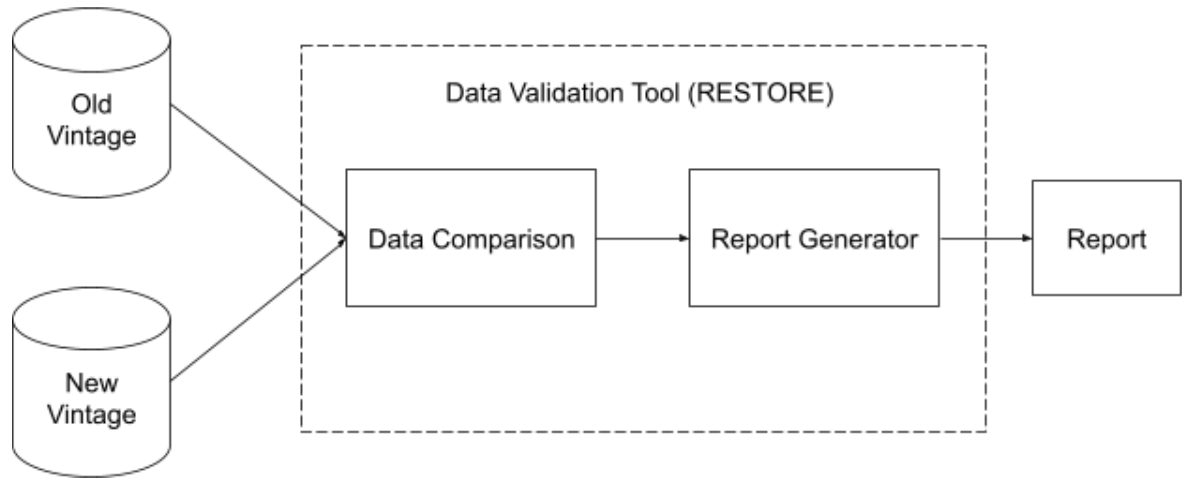}
    \caption{Comparison of old and new vintages of the dataset.}
    \label{fig:structure}
\end{figure}

These steps of automated data regression testing are graphically depicted in Figure~\ref{fig:structure}, which shows the following process (discussed in details in Sections~\ref{sec:tests} and \ref{sec:restore}).

\begin{enumerate}
    \item Load both old and new vintages of the dataset into RESTORE. 
    \item Apply a set of test to verify and validate the integrity of the new vintage.
    \item Generate the final report which is exported in human- and machine-readable formats. 
\end{enumerate}

Note that the report generated by RESTORE needs to be manually reviewed by the data scientists. The data scientists will use their experiences to decide whether a further investigation is necessary. The RESTORE package helps the data scientists to automate mundane data validation tasks and detect potential data defects as soon as possible (in the data preparation stage). We will discuss the assumptions and limitations of the proposed method in Section~\ref{sec:limitations}. Let us first look at the details of the tests used to compare the vintages.

\section{Tests' description}\label{sec:tests}

To address our problem (i.e., validation of the modified dataset), a set of tests performs an approximate comparison (rather than exact comparison as done by~\cite{DbFit}) of the datasets. Based on practical experiences, we create ten groups of tests, which will be discussed in details below. 

We also need to define success criteria, which consists of numeric thresholds, for these tests. The criteria are defined based on our practical experience, where we find that these values provide a large number of true data defects while keeping the number of false defects low. We cannot guarantee that these values are optimal for any dataset; rather, they can be treated as a set of good starting values and adjusted based on a particular use-case and the needs of a data tester.

Below we give details of our tests grouped into three categories: high-level tests dealing with metadata, tests of paired observations, and tests leveraging the results of the paired tests (which we deem higher-order tests). These tests are discussed in Sections \ref{sec:np-test}, \ref{sec:p-test}, and \ref{sec:h-test}, respectively. 

\subsection{High-level testing of vintages}\label{sec:np-test}
An example of a dataset vintage with $N$ variables\footnote{The term variable is synonymous to column or feature, depending on the reader's background.} and $M$ observations is given in Table~\ref{tab:vintage_ex}. Note that the `Key' values are not necessarily numeric. The only constraint is that the 2-tuple of `Key' and `Hierarchy Level' should be unique for every row (i.e., observation). 

\begin{table}
\caption{Example of a dataset vintage; $v_{\cdot}$ is a variable name. }
\label{tab:vintage_ex}
\centering
\begin{tabular}{r|l|r|r|r}
\toprule
Key      & Hierarchy Level & $v_1$    & $\ldots$ & $v_N$    \\ \midrule
1        & National        & 100      & $\ldots$ & 500      \\
2        & City            & 3        & $\ldots$ & 16       \\ 
$\ldots$ & $\ldots$        & $\ldots$ & $\ldots$ & $\ldots$ \\ 
$M$      & National        & 12       & $\ldots$ & 124      \\ \bottomrule
\end{tabular}
\end{table}

We now perform three groups of high-level tests assessing the characteristics of the vintages as follows:
\begin{enumerate*}
    \item comparing attributes of the variables, 
    \item checking the variables for missing observations, and
    \item counting discarded observations.
\end{enumerate*}

\subsubsection{Variables' attributes comparison}\label{sec:diff-test} 
\underline{Rationale}: We perform a set of ``sanity-check'' tests, comparing high-level characteristics of the datasets (i.e., metadata), such as the number of rows and columns. If the numbers do not match, this may be a cause for concern.\footnote{If machine learning techniques are used subsequently, data scientists may need to rebuild the machine learning model if they want to train the model on all the variables, but this is out of the scope of this paper.} Here, we assume that the datasets have no significant changes. Details of our assumptions can be found in Section~\ref{sec:limitations}. 

\underline{Method}: We obtain three items for each of the vintages, old and new: the number of variables, the names of variables, and the number of observations in the dataset. We then compare these items.

\underline{Success criteria}: If the number of variables, the names of the variables, and the number of observations are the same, then the test passes. If the number of variables or the number of observations is not identical between the old vintage and the new vintage of a dataset, the test fails and this mismatch gets reported. If a variable, present in the old vintage, is not present in the new vintage (or vice versa), then it is also considered a failure and the name of the variable is added to the report. 

\subsubsection{Missing (NA) observations}\label{sec:na-test}
\underline{Rationale}: Typically, a clean dataset should not have missing observations in a variable. 

\underline{Method}: Thus, for each `Variable Name' (e.g., for each $v_i$ in $v_1, \ldots, v_N$ in Table~\ref{tab:vintage_ex}) and `Hierarchy Level', we search for missing observations (in R such observations are marked as NA). This process is done individually for old and new vintages of the dataset.

\underline{Success criteria}: A `Variable Name' and `Hierarchy Level' pair that has zero missing observations passes the test; otherwise, it gets reported.

\subsubsection{Discarded observation count}\label{sec:disc_cnt}

Now we can join the old and new vintages of the dataset, so that we can perform pairwise tests for each variable (as will be discussed in Section~\ref{sec:p-test}). We perform the inner join (in the relational algebra sense of the term~\cite{maier_1983}) on the `Key' and `Hierarchy Level' columns (shown in Table~\ref{tab:vintage_ex}) of the old and new vintages, discarding the observations that are present in only one of the vintages. Before moving to the pairwise tests, we will perform one last metadata test, based on the count of discarded observations. 

\underline{Rationale}:  In practice, we found out that paired observations are more valuable for detecting defects than the non-paired ones (as they contain more information about changes to the dataset). However, the observations that did not make it into the inner join of the old and new vintage may indicate a defect in the data preparation process. 

\underline{Method}: Count the number of observations present in the old and absent in the new vintage, deemed $c_1$, as well as the number of observation present in the new and absent in the old vintage, deemed $c_2$. Note that we already compared the count of observations of the vintages in Section~\ref{sec:diff-test}. However, in this section we pair the observations, which brings additional information. If we denote  the count of observations in the old vintage as $c_o$, in the new vintage --- as $c_n$, and in the join of old and new vintage as $c_u$, then $c_1 = c_o - c_u$ and  $c_2 = c_n - c_u$. If all the observations are paired via the inner join, then $c_o = c_n = c_u$.

\underline{Success criteria}: The test passes if $c_1 = 0$ and $c_2 =0$; otherwise the test fails, and the values of $c_1$ and $c_2$ are reported.  A tester can then assess if the discarded observations appeared in the data are due to normal data churn or because of a defect in data preparation.

\subsection{Paired testing}\label{sec:p-test}
Once we join the old and new vintage (using the approach discussed in Section~\ref{sec:disc_cnt}), we can conduct the following sets of pairwise tests (performing comparisons for a given variable and hierarchy level):
\begin{enumerate*}
    \item the magnitude ratio test, 
    \item the mean relative error test, 
    \item the correlation test, and 
    \item the distribution test. 
\end{enumerate*}
These tests are discussed below.

\subsubsection{Magnitude ratios}\label{sec:magn}
\underline{Rationale}: We compare the magnitudes for the minimum, maximum, sum, mean, and median values for each level of hierarchy between the old and new vintages. The expectation is that extreme points of the distribution, as well as the central points, pairwise, should be in the same ballpark, which we will assess by comparing the order of magnitudes. 

\underline{Method}: Let us denote a metric for the $i$-th variable and $j$-th hierarchy level of an old vintage as $m_{i,j,o}$ and for $i$-th variable and $j$-th hierarchy level of a new vintage as $m_{i,j,n}$, respectively. Then the magnitude ratio $R_{i,j}$ is computed as follows:

\begin{equation}
	R_{i,j} = 
	\begin{cases}
		1, &\text{if $m_{i,j,o} = 0$ and $m_{i,j,n} = 0$;} \\
		\text{undefined}, &\text{if $m_{i,j,o} = 0$ or $m_{i,j,n} = 0$;} \\
		m_{i,j,o} / m_{i,j,n}, &\text{otherwise.}
	\end{cases}
\end{equation}

\underline{Success criteria}: We compute the value of $R_{i,j}$ for each pair of the metrics (min of the old and new vintage, max of the old and new vintage, etc.). If $0.1 < R_{i,j} < 10$ then both values are of the same magnitude and the test succeeds,\footnote{The threshold $R_{i,j}$ can be adjusted on a case-by-case basis, we will further in Section~\ref{sec:best-practices}.} otherwise --- fails and gets reported. Note that we have two special cases. If $m_{i,j,o} = 0$ and $m_{i,j,n} = 0$, then we assume that the magnitudes are identical --- setting $R_{i,j}=1$. If $m_{i,j,o} = 0$ or $m_{i,j,n} = 0$, then we cannot credibly assess magnitude difference; in this case we emit a warning asking an analyst to assess the magnitude difference manually. 

Note that given the pairwise nature of the comparison, the ratios of sums and averages will yield identical results. However, we retain both for a practical reason: the sums help an analyst to compare the values of variables at different levels of hierarchies (as, typically, the sum of observations at a lower hierarchy level aggregate to the value at a higher level of the hierarchy) hence the decision to keep the sum values.

\subsubsection{Mean relative error}\label{sec:mre}
\underline{Rationale}: The previous test (comparing min, max, etc.) assesses statistics that discard information about pairwise relations of individual observations. Given that we pair observations in the old and new vintage, we can compare each observation using mean relative error. We prefer the mean relative error over the mean absolute error because the values of attributes vary significantly between the variables as well as the variables' hierarchy levels.

\underline{Method}: Let us pair old and new observations for the $i$-th variable and denote paired vector of observations for the $i$-th variable and $j$-th hierarchy level of old vintage as $x_{i,j,o}$ and for the new vintage as $x_{i,j,n}$. Then the mean relative error $E_{i,j}$ is computed as an average of relative errors of each pair of observations in $x_{i,j,o}$ and $x_{i,j,n}$:
\begin{equation}
E_{i,j} = \langle \vert (x_{i,j,o} - x_{i,j,n}) \oslash x_{i,j,o} \vert \rangle, 
\end{equation}
for all non-zero elements of $x_{i,j,o}$, where $\oslash$ is the Hadamard division operator (performing element-wise division of vectors) and $\langle \cdot \rangle$ computes the mean.

By construction, all pairs of observation, where an element from $x_{i,j,o}$ is equal to $0$, have to be ignored. If $x_{i,j,o}$ vector has a lot of zero values, then this test may become misleading. In this case one can implement another test of relative change, see the work of~\cite{tornqvist1985should} for review and comparison of such tests.

\underline{Success criteria}: The test considered successful if $E_{i,j} < 0.2$.  A smaller value of the threshold can generate a high number of false alarms based on the our previous experiences.

\subsubsection{Correlation test}\label{sec:cor}
\underline{Rationale}: We expect that there should be a strong mutual relation between the observations of a given variable in the old and new vintages. To measure the strength of this relation, we compute correlations between the values of a given variable in the old and new vintages. The relation does not necessarily have to be linear but it should be monotonic. Thus, to assess these properties, we use Pearson product-moment correlation coefficient~\cite{pearson1896mathematical, myers_research_2010} (to assess linearity) and Spearman rank-order correlation~\cite{spearman1904proof, myers_research_2010}  (to assess monotonicity). 

\underline{Method}: We compute Pearson and Spearman correlations coefficients (deemed $r_{i,j}$ and $\rho_{i,j}$, respectively) for pairs of $x_{i,j,o}$ and $x_{i,j,n}$ for each variable $i$ and hierarchy level $j$. Correlation values range between $-1$ and $1$, with $1$ being perfect correlation, $-1$ --- perfect anticorrelation, and $0$ --- no correlation.

\underline{Success criteria}: The test is considered successful if $r_{i,j} \geq 0.8$ and $\rho_{i,j} \geq 0.8$, and unsuccessful otherwise (similar to the criteria in Section~\ref{sec:magn}, this threshold can also be adjusted on a case-by-case basis). From a practical perspective, a lot of real-world variables exhibit nonlinear relations (plus Pearson correlation assumes data normality which is often not the case). Thus, we pay more attention to the case of $\rho_{i,j} < 0.8$ than to the case of $r_{i,j} < 0.8$, because (empirically) they observed that it is a stronger indicator of a defect in the data.

\subsubsection{Distribution test}\label{sec:dist}
\underline{Rationale}: The previous test assesses the correlation between the $i$-th variable of the old and the new vintages. In this test, we generalize this approach by comparing distributions of the old and new vintages of this variable. If the distributions are significantly different, then it may be an indicator that there is a defect in the data. 

\underline{Method}: We use the nonparametric two-sample Kolmogorov-Smirnov test~\cite{smirnov1939estimation} to compare the differences between the two distributions. The null hypothesis of the test is that the samples are drawn from the same distribution. 

\underline{Success criteria}: The value of the Kolmogorov-Smirnov test $p$-value for the $i$-th variable and $j$-th hierarchy level is denoted by $S_{i,j}$. If $S_{i,j} < 0.05$, we assume that the null hypothesis is rejected and declare test failure. If $S_{i,j} \geq 0.05$ --- the test succeeds (even though it does not imply that the distributions are not different). 

\subsection{Higher-order testing}\label{sec:h-test}
The set of higher-order tests (i.e., those that combine the values of the metrics computed in Section~\ref{sec:p-test}) is composed of the following:
\begin{enumerate*}
    \item the comparison of Spearman correlation coefficients for different levels of hierarchy, 
    \item hybrid test, and 
    \item ranking of the number of test failures.
\end{enumerate*}
The details of the tests are given below.

\subsubsection{Comparison of Spearman correlation for different levels of hierarchy}\label{sec:pair-cor}
\underline{Rationale}: In Section~\ref{sec:cor}, we computed Spearman correlation $\rho_{i,j}$ for $i$-th variable and $j$-th level of hierarchy. Data scientists in our team observed that a significant difference in the $\rho$ values for two adjacent levels of hierarchy (i.e., $\rho_{i,j}$ and $\rho_{i,j+1}$) may indicate a defect in the data of the $i$-th variable. The root cause of such defect often relates to different aggregation procedures (from the raw data) associated with different levels of hierarchy. 

Note that while we compute both Pearson and Spearman correlations in Section~\ref{sec:cor}, the comparison test focuses only on the latter. As we discussed in Section~\ref{sec:cor}, the $\rho_{i,j} < 0.8$ (Spearman correlation) is a stronger indicator of a defect in the data than $r_{i,j} < 0.8$ (Pearson correlation). Analogously, it was found that comparison of differences in $\rho_{i,j}$ is a better indicator of a defect than a comparison of differences in $r_{i,j}$. Thus, to reduce tester's information overload, it was decided not to include the comparison of $r_{i,j}$ in the report. 

\underline{Method}: We compute relative difference $C_{i,j}$ between two adjacent levels of hierarchy:
\begin{equation}
	C_{i,j} = (\rho_{i,j} - \rho_{i,j+1}) / \rho_{i,j}, \text {if $\rho_{i,j} \neq 0$. }
\end{equation}
Given $J$ levels of hierarchy, with the $1$-st level being the top one and the $J$-th level being the bottom one, we perform $J-1$ computations of $C_{i,j}$, with $j = 1, \ldots, J-1$.

\underline{Success criteria}: Based on our experience, $ -0.1 < C_{i,j} < 0.1 $ is considered acceptable. $C_{i,j}$ values outside of this range may indicate a problem with the data of the $i$-th variable and $j$-th or `$j+1$'-th levels of the hierarchy. 

\subsubsection{Hybrid testing}\label{sec:hybrid-test}
\underline{Rationale}: We described multiple tests in the sections above. Intuitively, the higher the number of tests that failed for a given variable and hierarchy level $x_{i,j,n}$ --- the higher the chances that there is something wrong with the observations of this variable. We observed that a simultaneous failure of four tests --- namely, mean relative error (Section~\ref{sec:mre}), Spearman and Pearson correlations (Section~\ref{sec:cor}), and Kolmogorov-Smirnov test (Section~\ref{sec:dist}) --- is a very strong indicator of a defect in the underlying data (based on EA data scientists' past experiences). Thus, if $x_{i,j,n}$ fails all those tests, it should attract the attention of the data team.

\underline{Method}: We identify all the variables that failed four above-mentioned tests simultaneously and report them along with the values of the associated metrics. Table~\ref{tbl:mult_failure} displays an example of this report.  

\begin{table}[t]
\caption{Example: results obtained from the hybrid test. }
\label{tbl:mult_failure}
\centering
\begin{tabular}{l|l|r|r|r|r}
\toprule
Variable & Hierarchy & $E_{i,j}$ & $r_{i,j}$ & $\rho_{i,j}$ & $S_{i,j}$  \\
Name ($i$) & Level ($j$) &  &  &  &  \\ \midrule
$v_5$         & National        & 0.578               & 0.401               & -0.278                & 0.001 \\ 
$v_5$         & City            & 0.617               & 0.672               & 0.693                & 0.002                           \\ 
$v_8$         & City            & 0.669               & 0.532               & 0.454                & 0.046                           \\ 
\bottomrule
\end{tabular}
\end{table}

\underline{Success criteria}: As shown in Table~\ref{tbl:mult_failure}, a variable's name and corresponding hierarchies are listed in the report if and only if all of the following criteria are satisfied:
\begin{enumerate*}
    \item $E_{i,j} \geq 0.2$,
    \item $r_{i,j} < 0.8$,
    \item $\rho_{i,j} < 0.8$, and
    \item $S_{i,j}< 0.05$.
\end{enumerate*}

\subsubsection{Ranking of the number of test failures}\label{sec:ranking-test}
\underline{Rationale}: All of the above metrics are computed for each variable and hierarchy level individually. We observed that a test failure at multiple levels of the hierarchy of a given variable acts as a reliable indicator of a defect in the data associated with this variable. 

\underline{Method}: Thus, it is useful to count the number of test failure for each variable and test type and then order them in descending order from the highest number of test failures to the lowest. To reduce clutter, we report only the variables that have at least one test failure associated with them. Example of such ranking is given in Table~\ref{tab:tab2}. 

\underline{Success criteria}: An ultimate success is when there are no test failures associated with a variable and this variable does not show up in the report. The higher the number of tests and types of tests that failed --- the higher the chances that a variable has a defect in its data.

\begin{table}
\caption{Example: ranking test. To preserve space, a subset of tests is shown in this example.}
\label{tab:tab2}
\centering
\begin{tabular}{l|r|r|r|r|r}
\toprule
Variable Name ($i$) & $E_{i, \text{all}}$ & $r_{i, \text{all}}$ & $\rho_{i, \text{all}}$ & $S_{i, \text{all}}$ & Total  \\ \midrule
$v_7$    & 4   & 6        & 6       & 6            & 22  \\ 
$v_3$    & 4   & 5        & 2       & 5            & 16  \\ \bottomrule
\end{tabular}
\end{table}

\section{Discussion of tests' properties}\label{sec:discussion}
\subsection{Root causes of test cases' failures}
As mentioned at the beginning of Section~\ref{sec:tests}, not every test failure will lead to exposure of a data defect. Instead, a failure suggests that a new vintage is different from the old one in some unexpected way, and that a tester should take a close look at the failure. 
For the tests operating at a particular hierarchy level (i.e., those discussed in Sections~\ref{sec:na-test},~\ref{sec:p-test}, and~\ref{sec:hybrid-test}), a good starting point of an investigation is a review of data transformation procedures for a particular variable and level of hierarchy for which the test case failed. In the case of the test discussed in Section~\ref{sec:pair-cor} (examining adjacent levels of hierarchy), the problem typically is associated with data transformation procedures for one of these levels. The root cause of a failure of the test described in Section~\ref{sec:ranking-test} often resides in the general procedure that touches multiple levels of the hierarchy of the variable under investigation. 

The tests discussed in Section~\ref{sec:diff-test} operate at an even lower level of granularity (as they deal with potentially missing variables or observations). While removal or addition of variables is not uncommon, sometimes an analyst renames a variable by mistake, which often ends up being the root cause for the variable to appear in the report of this test. If the datasets have a significantly different number of observations, it may be caused by datasets truncation or data corruption. A failure of the final metadata-related test, discussed in Section \ref{sec:disc_cnt}, may indicate corruption of the values in the `Key' or `Hierarchy Level' columns.

\subsection{Predictive power of tests}
As discussed above, not every failure of a test ``translates'' into an actual defect. However, anecdotally, we observed that higher-order tests described in Sections~\ref{sec:hybrid-test} and~\ref{sec:ranking-test} yield the lowest number of false alerts~\footnote{We believe that these two tests will also help to mitigate the multiple comparisons problem~\cite{Gelman2012Why} because we are more interested in the observations that trigger multiple test failures in these two tests.}, followed by the correlation-related tests in Sections~\ref{sec:cor} and~\ref{sec:pair-cor}.  

On the other side of the spectrum, the distributions comparison test discussed in Section~\ref{sec:dist} yields the highest number of false alarms. This is expected, as the underlying distributions for a large number of variables in periodically updated datasets experience legitimate change to their underlying distributions (which the test detects successfully). However, the list of such variables are typically known to the dataset curators and, thus, can be filtered out with relative ease during the analysis of the report (generated by RESTORE). 

The rest of the tests fall in the middle of the spectrum. For example, the change to a distribution also translates into changes to statistics (such as mean, min, and max), which we analyze in Section~\ref{sec:magn}. However, because we are comparing the magnitudes of these statistics, these tests are less prone to false alarms. 

\subsection{Data types}\label{sec:data-type}
All of the tests can process variables to which ratio and, arguably, interval scales~\cite{stevens1946theory} can be applied. 

We will also be able to compute the test for numeric variables measured on nominal or ordinal scales~\cite{stevens1946theory}, but the results of some of these tests (e.g., magnitude comparison of averages for the ordinal scale) would be questionable from the statistical perspective. Thus, one has to be careful when interpreting the results of the tests.

The test cannot be computed for non-numeric tests, except for the tests discussed in Section~\ref{sec:np-test}. 

Fortunately, curators of datasets typically know data types and measuring scales of the variables in the datasets and can recommend which variables should be excluded from the analysis.

\subsection{Assumptions and limitations}\label{sec:limitations}
Here, we summarize assumptions on which our proposed method is based as well as associated limitations.
\begin{enumerate}
    \item As discussed in Section~\ref{sec:ea}, we assume that all input datasets for RESTORE are in tabular format (or can be transformed to this format). There exist other data formats, e.g., JSON and SQLite. We recommend practitioners to convert non-tabular format datasets into a tabular format before working with RESTORE.   
    \item We assume that there are no significant changes between two versions of the dataset. However, this might not always be the case. For example, in the case of the current pandemic, geodemographic data can change significantly in certain areas.\footnote{This may affect the downstream data science pipeline. For example, the subsequent machine learning models may have to be re-trained.}
    \item If there are attributes that only exist in one vintage dataset, we only report the names of such attributes. As discussed in the section of high-level testing (Section~\ref{sec:np-test}), our proposed test suite can report missing/new attributes by comparing all the names of attributes in both datasets. However, the tests discussed in Sections~\ref{sec:p-test} and~\ref{sec:h-test} are not applicable for missing/new attributes in such a circumstance.
    \item Our proposed method does not deal with categorical data. As discussed in Section~\ref{sec:data-type}, we focus on numeric tests. One can leverage our method for categorical data by converting it to numerical data using one-hot encoding~\cite{BeckW00}. However, this is beyond the scope of this paper.
    \item As discussed in Section~\ref{sec:mre}, zero values are ignored if they are denominators, because the mean relative errors will mathematically become undefined in such cases. Other formulas --- e.g., mean absolute percentage error (MAPE) --- may suffer from the same issue. If an analyst experience this issue, they may implement a metric designed to avoid the issue, such as msMAPE~\cite{chen2004assessing} or MASE~\cite{hyndman2006another}.
    \item The attributes that contain NA values are reported (as discussed in Section~\ref{sec:na-test}). Paired tests in Section~\ref{sec:p-test} are not applicable for observations with NA values. Data scientists can manually check the generated report and decide if an observation with NA values require further investigation.
\end{enumerate}

Note that one can create additional metrics for the test suite to address the above-mentioned limitations.

\section{The RESTORE package and take-away messages}\label{sec:restore}
In this section, we introduce the interface of the RESTORE package in Section~\ref{sec:interface}. Then, we validate RESTORE in industrial settings and find answers to our research question in Section~\ref{sec:validation}. After that, we discuss threats to validity in Section~\ref{sec:threats}. We also discuss potential extensions of RESTORE in Section~\ref{sec:potential}. In Section~\ref{sec:best-practices}, we deliver our take-away messages for practitioners applying data validation techniques.

\subsection{The interface of RESTORE}\label{sec:interface}
We implement the set of tests discussed in Section~\ref{sec:tests} in an open-source R package, available at~\cite{restore_repo}. The installation of the package follows a standard installation process for R package, details are given in the README file of~\cite{restore_repo}.
 
The tests are controlled by a single function \code{test\_two\_datasets}. The function ingests old and new vintages of the dataset as well as specification of the hierarchy either from CSV files or from R data frames. 
 
We found that for interactive testing, when a tester adjusted the datasets and wanted to quickly assess the results, the CSV files were more convenient. On the contrary, for automated testing, when the datasets were tested as part of the automated regression test harnesses, the data frame option was more suitable.
 
The final report is written into a user-specified XLSX file or saved as R data structure (so that it can be easily parsed later, if necessary). Users can select the test which should be stored in the final report. Parameters of the \code{test\_two\_datasets} function  are as follows.

The parameters \code{legacy\_file} and \code{target\_file} set the path to the files that contain the old vintage of the dataset and the new vintage of the dataset, respectively.
 
\begin{figure}
  \centering
  \subfloat[]{\label{fig:tree}
  \includegraphics[width=0.2\linewidth]{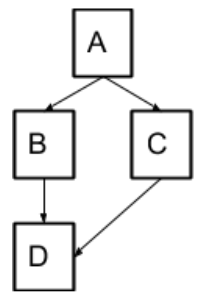}}%
  \enskip
  \subfloat[]{%
  \resizebox{0.36\linewidth}{!}{
    \begin{tabular}[b]{l|l}
    \toprule
    Parent & Child \\ Hierarchy & Hierarchy \\ Level & Level \\ 
    \midrule
    A & B \\ 
    A & C \\ 
    B & D \\ 
    C & D \\
    \bottomrule
    \end{tabular}
    }
    \label{tab:hierarchy_sample}
  }
  \quad
  \subfloat[]{%
  \resizebox{0.3\linewidth}{!}{
    \begin{tabular}[b]{r|l}
    \toprule
    Rank & Hierarchy \\ & Level \\ 
    & \\
    \midrule
    1     & A               \\ 
    2     & B               \\ 
    3     & D               \\ 
    4     & C               \\ 
    \bottomrule
    \end{tabular}
    }
    \label{tab:hierarchy_ranking_sample}
  }
  \caption{An example of hierarchy configuration. (a) Example of a non-linear hierarchy. (b) 2-tuple encoding of the non-linear hierarchy from Figure~\ref{fig:tree}. (c) Sample ranking of the hierarchy levels in Figure~\ref{fig:tree}.}
  \label{fig:figure-and-table}
\end{figure}

The parameters \code{hier\_pair} sets the path to a CSV file containing 2-tuples `Parent Hierarchy Level' and `Child Hierarchy Level', this enables RESTORE to operate on non-linear hierarchies. For example, a tree depicted in Figure~\ref{fig:tree} will be encoded by 2-tuples shown in Figure~\ref{tab:hierarchy_sample}.

The parameter \code{hier} points to a CSV file containing an ordered list of hierarchy levels, which is used for sorting the test results in the reports containing hierarchy column (e.g., the one shown in Table~\ref{tbl:mult_failure}), see Figure~\ref{tab:hierarchy_ranking_sample} for an example of such a file. Note that this parameter is not used to define the actual hierarchy.
 
The parameter \code{thresholds} points to a CSV file containing values for success criteria of tests described in Section~\ref{sec:tests}. 
 
The variables described above have corresponding ``twin'' parameters (namely,   \code{legacy\_df}, \code{target\_df}, \code{hier\_pair\_df}, \code{hier\_df}, and \code{thresholds\_df}) which allow to pass the dataset and configuration files in the R data frame format. 

The \code{final\_report} parameter specifies the location of the output report file in XLSX format. The parameter \code{final\_data} specifies an output location for the report stored in the R data structure format. The rest of the parameters are used to determine important variable names and a list of tests to run, as summarized in Table~\ref{tbl:params}. 
 
While RESTORE reads data only from CSV files or data frames, it does not imply that we cannot leverage other data formats. We simply need to convert our data into one of these two formats. For example, if the data resides in a relational database, one can issue SELECT SQL query from R using \code{DBI}~\cite{dbi} package, which will automatically extract and convert the data into the R data frame format. 

As part of the package, we provide a sample file demonstrating the usage of RESTORE program interface (see \code{example.R} on Zenodo~\cite{restore_repo}).

\begin{table*}[!th]
\centering
\caption{Additional parameters of \code{test\_two\_datasets}.}
\resizebox{\linewidth}{!}{
\begin{tabular}{l|l|l|l}
\toprule
Parameter & Type & Details & Default Value 
\\ \midrule
\code{key\_col} & String & Name of the column containing `Keys'. & NA \\ \hline
\code{hier\_col} & String & Name of the column containing `Hierarchy Levels'. & NA \\ \hline
\code{report\_char} & Boolean & TRUE --- generate the report of characteristic comparison; FALSE --- do not generate the report. & TRUE \\ \hline
\code{report\_na} & Boolean & TRUE --- generate the report of missing (NA) variables; FALSE --- do not generate the report. & TRUE \\ \hline
\code{report\_discard} & Boolean & TRUE --- generate the report of discarded observations; FALSE --- do not generate the report. & TRUE \\ \hline
\code{report\_magnitude} & Boolean & TRUE --- generate the report of magnitude metrics; FALSE --- do not generate the report. & TRUE \\ \hline
\code{report\_mre} & Boolean & TRUE --- generate the report of mean relative errors; FALSE --- do not generate the report. & TRUE \\ \hline
\code{report\_spearman} & Boolean & TRUE --- generate the report of Spearman test; FALSE --- do not generate the report. & TRUE \\ \hline
\code{report\_pearson} & Boolean & TRUE --- generate the report of Pearson test; FALSE --- do not generate the report. & TRUE \\ \hline
\code{report\_distribution} & Boolean & TRUE --- generate the report of distribution test; FALSE --- do not generate the report. & TRUE \\ \hline
\code{report\_spearman\_diff} & Boolean & TRUE --- generate the report of Spearman comparison for hierarchies; FALSE --- do not generate the report. & TRUE \\ \hline
\code{report\_hybrid} & Boolean & TRUE --- generate the report of hybrid metrics; FALSE --- do not generate the report. & TRUE \\ \hline
\code{report\_ranking} & Boolean & TRUE --- generate the report of ranking; FALSE --- do not generate the report. & TRUE \\ \bottomrule
\end{tabular}}
\label{tbl:params}
\end{table*}

\subsubsection{Special case: flat hierarchy}\label{sec:special}
To deal with the case of a flat hierarchy (i.e., non-hierarchical dataset), we do not need to pass \code{hier\_pair} and \code{hier\_pair\_df} values to \code{test\_two\_datasets}. Under the hood, RESTORE adds a dummy hierarchy column to the dataset and runs all the tests against the dataset except for the test comparing correlation coefficients for different values of hierarchy (discussed in Section~\ref{sec:pair-cor}).

\subsection{Validation}\label{sec:validation}

The RESTORE R package has been institutionalized into EA's product development cycle. The data scientists use the package to detect defects in the new vintage of the datasets. To assess the benefits of the package, we seek an answer to the following question.

\textbf{Does the RESTORE package improve the efficiency of the data validation procedure, i.e., identifying data errors with less time and human resources?}

First, we quantified the time needed to run all the tests on two reference geodemographic datasets (named $D_1$ and $D_2$). The summary statistics for these datasets are shown in Table~\ref{tbl:perf}. The table also shows the average and the standard deviation of the execution time of \code{test\_two\_datasets} function based on 10 runs of the function for each dataset. We kept the parameter values of the function to defaults, i.e., all of the reports were generated.

Potential data defects have been successfully detected and reported by RESTORE (note that whether a result reported by RESTORE is a true data defect will require manual investigation by data scientists). To preserve space, we will highlight two examples. The examples can also be found in the demo uploaded to Zenodo~\cite{restore_repo}. The first example depicts how Spearman rank-order correlation will list any variable names (as well as all the test results of these two variables, e.g., Pearson correlation) as long as the test results are below the success criteria (set at 0.8). The first example lists five variables (e.g., one of them has a Spearman correlation 0.6607) recorded in the output file. The second example shows a case in the demo where the mean relative error is bigger than the threshold of 0.5, which is 1.2342. There are 20 cases listed in the test of the mean relative error. All of them need to be manually examined by the data scientists. 

Our testbed is a laptop equipped with 2 GHz Intel Core i5 CPU and 16 GB memory, running R v.3.5.1 on MacOS v.10.14.3. The datasets are read from files (which is slower than reading the datasets from R data frames). Executing a complete set of tests and generating the final report took, on average, $\approx 3.4$ minutes for the $D_1$ and $\approx 2.3$ minutes for the $D_2$. 

The data scientists also found these tests are computationally inexpensive. 
Based on the feedback from EA data scientists, the same set of tests, when conducted manually by an experienced data tester takes $\approx 2$ hours of the tester's time (per dataset). Thus, using RESTORE speeds up\footnote{Note that we do not take into account the analysis of the test results. However, this time would be identical for both manual-based and RESTORE-based workflows.} this testing process by $\approx 97\%$.

\begin{table}[t]
\centering
\caption{Performance evaluation of RESTORE on two reference datasets. Note that the new vintage of $D_2$ consists of fewer variables compared to the old vintage. Consequently, less computations for comparison are needed --- hence less time compared to the one with $D_1$.  }
\label{tbl:perf}
\begin{tabular}{l|r|r|r|r|r|l}
\toprule
Dataset & \multicolumn{1}{c|}{Hierar.} & \multicolumn{2}{c|}{Variables} & \multicolumn{2}{c|}{Observations} & \multicolumn{1}{c}{Time}         \\
\multicolumn{1}{c|}{Name}                       & \multicolumn{1}{c|}{Levels}  & \multicolumn{2}{c|}{Count}     & \multicolumn{2}{c|}{Count}        & \multicolumn{1}{c}{$\pm$ St. Dev.}   \\
\cmidrule{3-4} \cmidrule{5-6}
                      & \multicolumn{1}{c|}{Count}   & \multicolumn{1}{c|}{Old}           & \multicolumn{1}{c|}{New}           & \multicolumn{1}{c|}{Old}             & \multicolumn{1}{c|}{New}            & \multicolumn{1}{c}{(Seconds)}   \\  \midrule
$D_1$ & 7 & 757 & 762 & 67,370 & 67,370 & $202 \pm 11$ \\ 
$D_2$ & 7 & 716 & 584 & 67,370 & 67,370 & $137 \pm 8$ \\ \bottomrule
\end{tabular}
\end{table}

The usefulness of RESTORE is also supported by results of an anonymous survey of 15 data scientists in EA who are now using the tool. An anonymous poll was sent to everyone with the following three questions.

\begin{enumerate}
    \item Is RESTORE helpful? Possible answers were ``Extremely useful'', ``Very useful'', ``Somewhat useful'', ``Not so useful'', and ``Not at all useful''.
    \item Does RESTORE save time? Possible answers were ``Yes'' or ``No''.
    \item Does RESTORE identify errors? Possible answers were ``Yes'' or ``No''.
\end{enumerate}

One respondent ($\approx 7\%$ of respondents) found RESTORE extremely useful, eleven respondents ($\approx 73\%$ of respondents) --- very useful, and three respondents ($20\%$ of respondents) --- somewhat useful. All fifteen respondents unanimously agreed that RESTORE saves time and identifies errors. The results suggest that the data scientists in EA find RESTORE to be a helpful data validation tool in the data development process because it can efficiently identify data errors. 

We have seen the results showing increased efficiency in the identification of potential errors in updated datasets with RESTORE. How does EA benefit from this efficiency in production?
RESTORE has been institutionalized by EA and integrated into the dataset development process discussed in Section~\ref{sec:ea}. Incremental changes made to the new vintage of the dataset are tested by RESTORE to make sure that the new vintage did not regress. If the tests failed, a root cause detection of the regression is easy to detect, as the failure is typically related to data transformations applied between the two increments.

In the long term, this increased automation and a better approach to data testing will significantly reduce the cost of delivering products to market. We estimate that the time of getting data to market can be reduced by about half (leading to cost reduction and improving customers' satisfaction).

To summarize, the data scientists found these tests to be helpful in practice, i.e., the tests could detect defects in the data reliably. The tests are also computationally inexpensive, which helps to preserve scalability and enable fast verification that the latest changes to a dataset did not inject any new errors. 

\subsection{Threats to validity}\label{sec:threats}
The threats to validity, classified as per \cite{wohlin2012experimentation,yin2009case}, are discussed below.

\subsubsection{Construct validity}
To assess the usefulness of RESTORE, we conducted the survey in EA. We choose an anonymous poll to proactively address the concern that EA data scientists can be positively biased toward the tool that is developed with the help of people from the same organization. We cannot formally prove that the questionnaire is sufficiently detailed to observe and measure bias-inducing factors (e.g., the work culture and politics) that may affect the results of the survey. However, the fact that EA institutionalized RESTORE implicitly supports the results of the survey.

\subsubsection{Internal validity}
The implementation of RESTORE that we wrote in R may contain bugs. To mitigate this threat, we performed peer code reviews and wrote automated unit tests. EA data scientists performed acceptance testing. 

\subsubsection{External validity}
We cannot claim that our software would be of use for any dataset, which is in line with other software engineering studies,  suffering from the variability of the real world~\cite{26c990771bd645428c33ea107259ceb5}.  However, RESTORE is helpful for structured hierarchical and non-hierarchical datasets, where at least some of the attributes exist in multiple vintages. These attributes should be either numeric or should be convertible to numeric values (as discussed in Section~\ref{sec:limitations}). 

Moreover, we cannot claim that our list of metrics is exhaustive. To mitigate this threat, we published our source code (as well as a demo program) on GitHub so that anyone can adopt and extend RESTORE for their needs.

\subsection{Potential extensions of RESTORE}\label{sec:potential}
The current version of RESTORE works by focusing on a pair-wise comparison of numerical datasets (measured using ratio and interval scale, as discussed in Section~\ref{sec:data-type}) that can be loaded into memory. This is sufficient for our use-cases. We released RESTORE as an open-source package so that one can extend or alter the tests implemented in RESTORE based on their specific use-cases or requirements. Below, we sketch potential ways to extend the package if one needs to compare large volumes of data, desires to compare other types of variables, or would like to do non-paired comparison of variables. 

RESTORE has been validated in production to process medium size datasets (e.g., datasets that comprise 600+ variables with $\approx$~1.5 million observations). Currently, RESTORE reads all data into memory. This may be an issue for very large datasets (a.k.a. big data). This can be mitigated by altering the process of ingestion datasets into the package: rather than loading the whole dataset into memory, one can process a subset of columns (e.g., loaded using \code{fread} function from R \code{data.table} package~\cite{datatable}) in multiple iterations.\footnote{Given that computations for every variable are independent of each other, the computations can be easily parallelized using \code{foreach}~\cite{foreach} and \code{parallel}~\cite{r_core} packages.} Alternatively, if the number of observations is such that they cannot be loaded into memory, then one can leverage an external framework, such as Spark, and perform the computations outside of the R engine. Note that Spark integrates into R, e.g., using \code{sparklyr} package~\cite{sparklyr}.

If a tester needs to apply RESTORE to other types of data, some of the tests (discussed in Section~\ref{sec:data-type}) are readily applicable. One can extend the package by adding additional tests. For example, to extend comparison of distributions to ordinal data, one can adopt Mann-Whitney U test~\cite{mann1947test}. 

Finally, as discussed in Section~\ref{sec:disc_cnt}, we did not perform comparison on non-paired observations (i.e., non-joined ones) of the datasets, as, empirically, they were found less useful for detecting defects in our datasets. However, if one desires to apply the tests to non-paired observations of a given variable, then it can be done with relative ease --- all the tests, with the exception of the ones discussed in Sections~\ref{sec:mre},~\ref{sec:cor}, and~\ref{sec:pair-cor}, are applicable to non-paired data.

\subsection{Take-away messages (the best practices)}\label{sec:best-practices}

Based on our hands-on experiences, we introduce five suggestions for data scientists who will develop data regression testing tools. To the best of our knowledge, our proposed best practices are the first general guidelines proposed for data scientists who want to adopt automated data validation in data preparation. In general, a data validation framework for datasets should have five characteristics as follows, and data scientists should adopt best practices to incorporate in the data validation these characters.
\begin{enumerate}
    \item \textbf{Comprehensive test oracles}. Data scientists should have explicit knowledge of the data structure and the reasonable ranges of values in the datasets. The testing framework will include multiple testing rules to test various aspects that may negatively affect the data quality.
    \item \textbf{Automated testing}. Compared to manual testing, where a set of tests is performed manually, automated testing significantly reduces the human and time resources for the testing process. 
    \item \textbf{Modularity-driven testing}. Keep in mind the variability of datasets. Thus, there is no one-size-fits-all solution. A scalable testing framework can be altered based on the nature of datasets. The set of tests can be truncated or extended so that all the erroneous data are discovered, and the false alarm rate is kept within a reasonable range (whichever data scientists are comfortable with).
    \item \textbf{Flexible thresholds}. The thresholds in testing can be set and adjusted based on empirical studies. For example, data scientists can decrease the threshold if the sensitivity is high. Thus, it is essential to keep the thresholds adjustable in the design of the testing framework. In this way, data scientists can always change the testing criterion if needed.
    \item \textbf{Continuous integration}. In case there are more than one data scientist who is responsible for data testing, or there are more than one data development teams, we would better use the best practices to manage code dependencies. For example, a version control system, such as git~\cite{spinellis2012git}, can be employed to empower people to collaborate. Besides that, we should also aim to ensure the testing framework to run on different environments, e.g., Linux and Mac OS. In such a manner, it is easy to share between team members (or different teams) without complicated configuration.
\end{enumerate}

Our experience can be extended to a more general software testing life cycle (STLC). There are various processes of STLC, they all must include, in some form, the four fundamental activities~\cite{hooda2015software}, i.e., 1)~requirement analysis and test analysis, 2)~test planning and preparation, 3)~test case development and test execution, and 4)~test cycle closure. The stages of STLC are activities conducted during the test procedure. Our experience report focuses on some best practices for those who want to adopt automated testing. Those best practices can be performed at any stage of STLC, e.g., automated testing can involve requirement analysis and testing planning. 

Note that our data regression testing take-away messages resonate with some best practices of software regression testing in software. However, the former focuses on data defects while the latter investigates software defects. We argue that both, data and software regression testing, are crucial to ensure the overall quality of data-driven software.

\section{Related work}\label{sec:related}

\emph{Software engineering best practices for data science}. Jones~\cite{jones2009software} gives an overview of software engineering best practices and introduces 50 best practices based on the study of long lifecycle projects in the industry. Kim et al.~\cite{Kim0DB16} discuss software-oriented data analytics. They conduct a survey on data scientists in Microsoft and propose a set of strategies to help data scientists increase the impact and actionability of their work. Begel and Zimmermann~\cite{BegelZ14} present the results from two surveys related to data science in the field of software engineering. The results list the most concerned questions from practitioners in 12 categories, including best practices of development and testing practices. 

\emph{Data validation frameworks}. \cite{gao2016big} provided a survey of big data challenges and data validation functionalities in existing big data platforms, such as Microsoft Azure HDInsight~\cite{CloudCom36:online}. While there exist various data validation tools, most of them focus on testing a single dataset (e.g.,~\cite{GreatExp24}) rather than comparison of two datasets, which is a key difference between our solution and other data validation tools. Comparing Great Expectations with RESTORE, Great Expectations is not designed for hierarchical data and thus is not suitable for proper testing of geodemographic data. IBM SPSS Statistics~\cite{ibmspss} can identify suspicious and invalid cases, variables, and data values in a given dataset. IBM SPSS Statistics focuses on providing a set of data validation rules for a single dataset, e.g., flag incomplete IDs. However, it cannot validate two versions of the same dataset. Thus, this approach is complementary to ours, which automates the checks of two datasets, saving an analyst's time. In the SPSS case, such checks would have to be built from scratch. \cite{bonter2012data} developed a data validation protocol for a bird monitoring program. The tool validates and filters the data to remove potential sources of errors and bias. \cite{sadiq2004data} identified the importance of data flow validation in workflow processes. They defined some possible errors in the data flow, e.g., missing data and redundant data. \cite{polyzotis2017data} investigated four common data management challenges (data validation is one of them) in machine learning. Later, this research group in Google~\cite{polyzotis2019data} studied data validation in the machine learning pipeline and identified the importance of detecting data defects early because the model trained with buggy data often amplifies the data bugs over a feedback loop. They proposed a data validation to quantify the distribution distance between training data and new data. However, automated tests are not in the scope of this work. Hence, it is hard to merge their methods or tools into a continuous integration pipeline. The data linter~\cite{hynes2017data} focused on data validation in the ETL pipeline. It is a machine learning-based tool that can automatically inspect input datasets for machine learning models, identify potential data issues (lints), and suggest potentially useful feature transforms for a given model type. The data linter does not compare two versions of datasets for data validation.

There exist some work to ensure data integrity during data migration~\cite{paygude2013automated,paygude2013automation,rathika2014automated}. Instead of comparing two datasets, their approaches detect if the data is altered during the migration process. Thus, we cannot apply their approaches in our case. Below, we also provide some other related but complementary papers.

\emph{Database testing frameworks}. There exists a significant amount of test frameworks for testing database engines and business logic that alters the data in the databases~\cite{kapfhammer2003family,maule2008impact,haraty2002regression,nanda2011regression,haftmann2005efficient}. In addition, some database testing frameworks are available~\cite{DbFit,DbUnit,NDbUnit,DBTest,SQL}. However, none of them are suitable for testing dataset vintages. 
 
\emph{Open source projects for regression testing of databases}. Regression testing tools for databases try to assure that a query (captured in one of the previous releases) executes successfully (in the release under test). This functionality is available in many existing automated database testing frameworks~\cite{DbFit,DbUnit,NDbUnit,DBTest,SQL}. However, this will typically be inadequate for our needs as successful execution of a statement cannot guarantee that the returned results are correct (as was discussed in Section~\ref{sec:intro}). Some database testing frameworks, e.g.,~\cite{DbFit}, can readily check if the recordsets are identical and highlight the difference between them. However, as we discussed before, changes between vintages of a dataset are expected. Thus, these tests are not sufficient for our needs.

\section{Conclusion}\label{sec:conclusion}

In this paper, we focused on applying software engineering best practices to the automation of data validation. Our data under study is taken from the geodemographic domain. We presented a set of tests that enable automated detection of defects in a new vintage of a dataset. We implemented the tests in an open-source R package called RESTORE and validated it in practice. We showed that the adoption of RESTORE can help the procedure of data validation achieve 
\begin{enumerate*}
    \item efficiency --- reducing the cost to incorporate a new vintage of a dataset,
    \item simplicity --- encapsulating a batch of relatively complex testing rules into one interface, and 
    \item scalability --- processing datasets comprised of about 1.5 million observations and more than 600 variables. 
\end{enumerate*}    
Moreover, we also proposed a set of strategies for the best practices in data validation based on our own experience.

This set of tests is of interest to practitioners, as using the RESTORE package on their datasets gives them the advantages to 
\begin{enumerate*}
    \item have more certainty about delivery dates for products,
    \item reduce the occurrence of data defects in products, and
    \item dedicate more time to developing new functionality, rather than testing the existing one.
\end{enumerate*}

\section*{Acknowledgment}
The work reported in this paper is supported and funded by Natural Sciences and Engineering Research Council of Canada, Ontario Centres of Excellence, and Environics Analytics. We thank Environics Analytics data scientists for their valuable feedback.

\bibliographystyle{abbrv}
\bibliography{reference}

\end{document}